**The emergence of a new sound research methodology in the field of health: *designo-therapy* ?**

1. Luc Perera, Phd student, EnsAD UNIVERSITY PSL , MINES ParisTech, Paris, France, luc.perera@cyu.fr
2. Pierre Jouvelot, Senior Research Scientist atMINES ParisTech UNIVERSITY PSL, Fontainebleau, France, pierre.jouvelot@mines-paristech.fr

Design fits in different fields, and it is given a plurality of titles: thinking, social, ecological, and graphic, space, etc. This discipline crosses the fields of research and engineering, which emancipate themselves from their historic fields and target other areas of public and private services. On the other hand, the sound sector, and more generally the acoustics, is more strictly categorized: musicology, psycho-acoustics or electro acoustics (see figure 1, which gives a relatively comprehensive overview) ). It is in this universe that evolves the discipline of sound design, with accents a priori industrial or environmental. But could it also allies, more unexpectedly, to health design and, if so, in what form ? To answer this question, we try here, in a few lines, to explain the links between art (s) and health, links that aroused the interest to draw a parallel with the world of design, then to think about its integration. In the medical community. We will conclude, as an illustration, from our own research in sound design, in connection with Indian music.

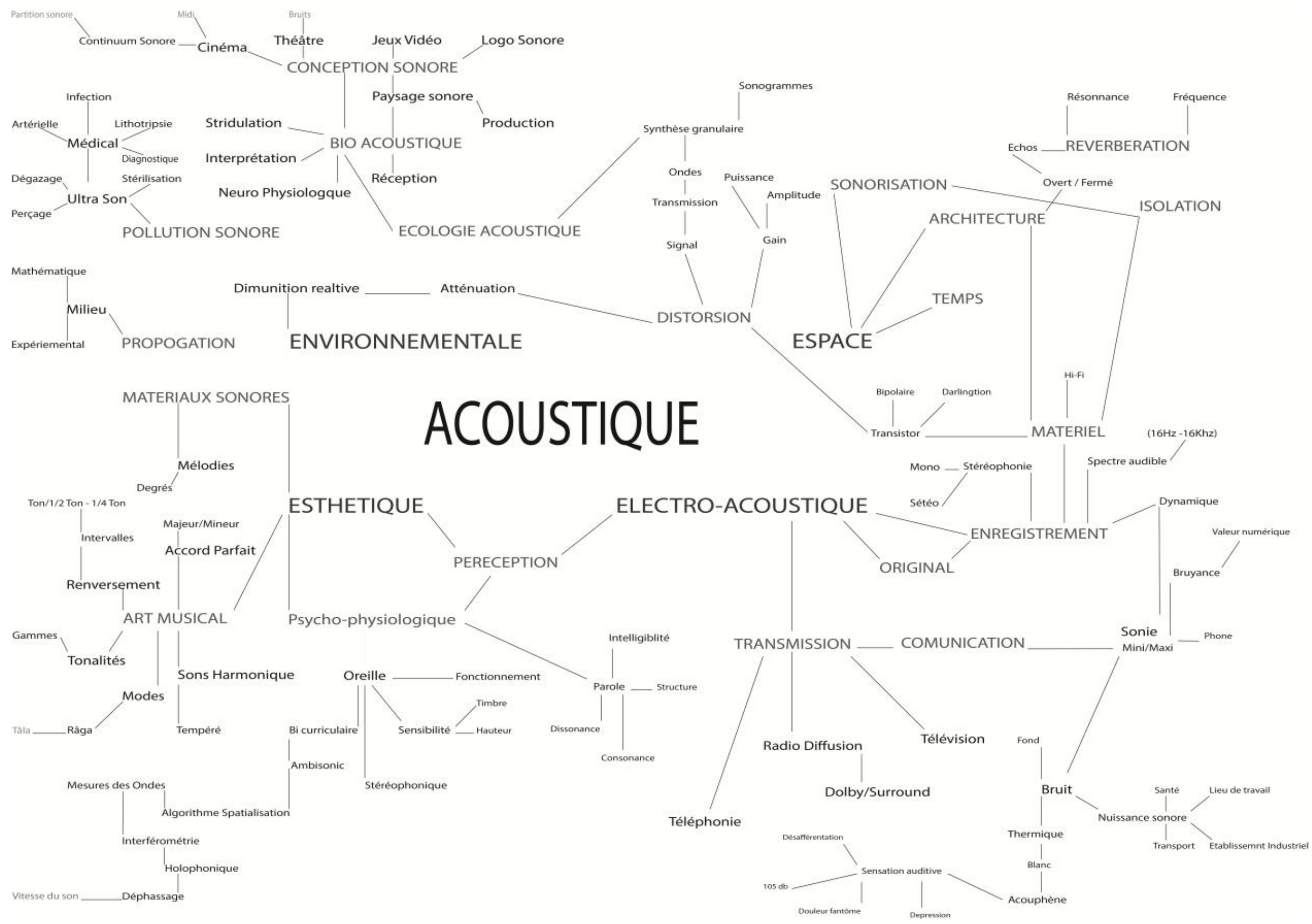

*Figure 1: Mind map of the word Acoustics, January 2019.*

## 1.1 Art Therapies

The links between art (s) and health in the modern sense have existed in hospitals and medical centers since the 1980s. This is observed (Bubien, 2004): "thanks to culture, the patient is considered in his globally, she can help her in her healing process. " They can even be quantifiable; what is termed therapeutic observance (Haynes et al, 1978) is a method that makes it possible to verify, from numerical data such as blood pressure, whether or not there is a benefit of an artistic practice. Art therapy is also a complement to non-drug management, this practice sometimes helping to relieve or recover abilities, through the only artistic practice as the city (Brioulet, 2016):
"The art-therapeutic work oriented towards the restoration of the executive functions, even if it did not allow an effective improvement of the executive functions out of session, contributes to the maintenance of a relational commitment in the project of care of the patient."

Moreover, in recent years, there has been an increase in art-based therapies in France, as noted, for example, (Oland, 2016), in the field of mental health: "In recent years, new Brain stimulation techniques have emerged in psychiatry. These are non-invasive and painless techniques. We can assume that patients see this as an improvement in their well-being and use it as a substitute in their daily lives. Many other artistic approaches are used in the medical community, even if they are not always recognized by the profession. A striking case is that of music therapy, which has existed since the 1960s in France. "In the years 1960-1970, a first French center of music therapy was created, in Paris, under the name Research Association of this application of psychomusical techniques (ARTP). (Vrait, 2018). The therapeutic method consists here of using reminiscence by passive listening: the patient listens to the music, and that awakens in him / her memories. There is also a so-called active variant: the patient then participates in a choir or practices an instrument. The study of (Chevrau, 2016) notes, in the field of dementia:

"The musical memory is resistant enough to the cognitive deterioration observed in neurodegenerative pathologies. By musical memory we mean the capacities related to the production of music (the practice of an instrument, the singing of an air, a song) and the semantic knowledge related to musical works ".

This musical practice is thus widely used in all that is related to neurodegenerative problems. Thus (Platel, 2011) even remarks:

"In the field of neurodegenerative diseases, music is an interesting medium in the regulation of mood, but seems above all to be a cognitive" stimulator "that has allowed us to reveal preserved abilities of implicit learning up to an advanced stage. of pathology. "

In recent years, the field of design has been able to integrate this new field of health, because its approach to the social sciences resonates in the medical environment, as suggested (Ruby, 2011) in his study of the relations between design and science:

"At first glance, and from a rapid perspective, this question has nowadays become that of design, that of an articulation into a single thing of an aesthetic, technical aspect (the usefulness and the formalization of human gestures) and sometimes scientific. "

The design tries to be in connection with the research and problems resulting from hospital constraints. The journal *Sciences du Design* n ° 6 offers case studies illustrating the role of design in the context of patient experience, aging, participatory design, etc. There is a proposal to define a methodology *by design* (Findeli, 2015) in the context of issues arising from the daily life of a hospital or medical center

## 1.2 Sound design and health

An often neglected and yet systemic aspect in the field of health is that of the sound environment of the medical community. In this field, sound design was at first simply limited to taking into account acoustic or psychoacoustic phenomena such as neonatal sound processing. As noted (Khun, 2016, p.167), since "any neuronal activation after sensory stimulation is accompanied by an increase in oxygen consumption that is measurable in NIRS[1]", the quality of the sound environment also had to be to be controlled. Then we became aware of the benefits of noise reduction in hospital rooms, because patients often kept their TV turned on at high volume, as found in the study (Leroux, 2006, p.22):

"The background sound of the early afternoon also plays on the atmosphere of the space concerned. In many services, the television set is on and / or a musical background is broadcast. The volume and choice of television program or type of music varies depending on the choice of caregivers rather than residents. "

[1] *Near Infrared Spectroscopy*

Beyond the music therapy, the sound was thus for a long time only a secondary element in the search for the good of the patients. For three years in France, schools or institutions such as ENSCI, Sciences Po, Ircam, TALM, EnsAD and MINES ParisTech try to meet the needs of patients with sound design. Seminars and symposia punctuate this awareness. Thus, even in 2018, we can note the seminar "*Design in health: how to change scale*[2]", where different designers, architects and carers came to discuss issues related to the hospital as the space layout to take care, therapeutic objects and health technology, but also issues around the sound. During the symposium organized, still in 2018, at Le Mans "Art du so (i) n:, sons, espaces et à l'hôpital"[3], master students presented their various projects and research in the fields of sound; the debates, which were very interesting and informative, discussed in particular the delicate question of the research protocols set up in the Ehpad. The 8th symposium on health will take place in Marseille. Health has thus become a means of capitalization and profitability of the Silver Economy (Numa, 2015). For exemple,

"The private equity firm Innovation Capital announced on February 25, 2014 the launch of the Sector Fund of the sector of the Silver Economy dedicated to funding Innovative Services for the actors of health and autonomy (SISA). SISA will start its first closing for more than 40 million euros. "

And the proliferation of FabLabs in medical centers only reinforces the link between these two worlds, as described (Fabbri, 2016):

"Alexandra Le Chaffotec opens the section devoted to specialized spaces on the case of a living lab dedicated to innovation in health. She finds that the space and time units specific to this type of space stimulate and enhance the innovation process by placing the patient at the center of the innovation process. "

## 1.3 Towards the Tala Sound

It is by taking into account this existing in the medical fields, music therapy and sound design that our research, rooted in what we do not hesitate to call *designo-therapy*, tries to encompass and make coexist their interdisciplinarity , despite its difficulties (Gentes, 2015, p.103):

"The difficulty in grasping it is therefore linked to its very status as inter-discipline and to the fact that the disciplines it uses, whether it be the hard sciences, the life sciences, the social sciences or the humanities, work with different disciplines. modalities and use different concepts. "

Thus, we did not hesitate to assert a bias (Ponge, 1967) sound in ethnomusicology: that of Indian music. In addition to the fact that this type of language is poorly studied in France, one of the authors, originally from Pondicherry, in the south of India, wanted to develop a device that would use Carnatic rhythms and that would be in opposition to the methods of listening "classic" (in the social sense) in hospitals. The objective thus becomes twofold: first, to see if this music influences the behavior of the patients by changing their listening mode (Author, 2018) and, secondly, to integrate into this multidisciplinary approach a methodology that could be inspired by social design (Gauthier et al., 2015):

---

[2] Seminar organized by the Sciences Po Chair of Health from 12 to 15 January 2018 at the Paris Diderot University.

3 Organized by the School of Art and Design of Le Mans and IRCAM, on December 6, 2018 at Le Mans, in collaboration with the Hospital Center, the Victor Hugo Clinic and the Jean Bernard Center and the support of Mans Acoustique and Mans Innovation

"Therefore, any design practice is necessarily social, in the sense that one of its fundamental problems is to implement a social and philosophical anthropology of the appreciation of ordinary life in the world. life with objects, places, services, institutions and organizations. "

While this approach draws inspiration from design and design research, it is also inspired by medical approaches, for example by leading to the development and testing of research protocols, in this case at the hospital. Paul Brousse in Villejuif (94) and the center Apad (Establishment : Accommodation for People Aged Dependents) in Verriere (78) France. The final objective is to recover metadata and to develop a specification for the design of a sound device, Tala Sound, which will react in relation to the Carnatic rhythmic characteristics. This musical structure may be codified in the form of a circle ( Figure 2) as a cycle of time in a continuous phase and the modification of which will be carried out via a computer system. Such an object will be used by a music therapist or nursing staff to calm the anxiety potentially related to his state of dementia. It will be driven from a simple software in which it will be possible to define, for example, the volume, the tempo and the choice of songs. If Tala Sound proves effective, we will have rallied these different disciplines of health, sound and design around a common project for the well-being of patients and hospital staff.

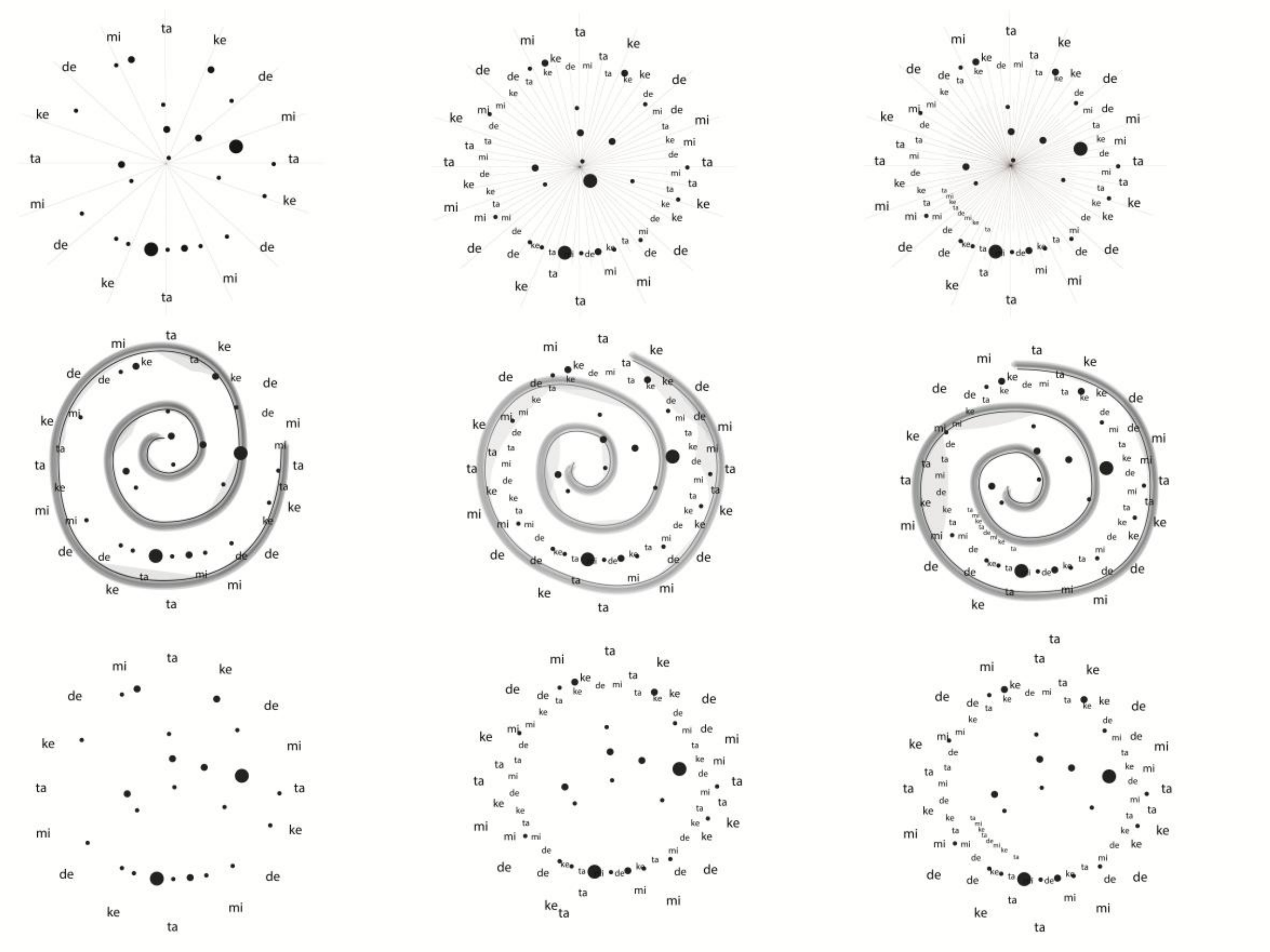

*Figure 2: Cycle change in Carnatic rhythm, January 2019.*